\documentclass[%
 reprint,
 amsmath,amssymb, aps
]{revtex4-2}

\usepackage{graphicx}
\usepackage{dcolumn}
\usepackage{hyperref}
\usepackage[normalem]{ulem}
\usepackage{xcolor}
\usepackage{mathrsfs}
\usepackage{cancel}
\usepackage{bm}
\usepackage{tikz-cd}
\usepackage{xtab,afterpage}
\usepackage{bbold}
\usepackage{leftidx}
\usepackage{upgreek}

\hypersetup{
    unicode=true,           
    pdftoolbar=false,        
    pdfmenubar=false,        
    pdffitwindow=false,      
    pdfstartview={FitH},     
    colorlinks=true,         
    citecolor=blue,           
    linkcolor=blue,
    urlcolor = blue
}
\usepackage{braket}
\usepackage{nicematrix}

\usepackage[T1,T2A]{fontenc}
\usepackage[utf8]{inputenc}
\usepackage[english]{babel}
\usepackage{mathtools}

\begin{document}

\title{Photon-mediated entanglement between spin qubits beyond the dispersive regime}

\author{\firstname{Andrei I.}~\surname{Nikitchenko}}
 \email{andrei.nikitchenko@uni-konstanz.de}
\affiliation{Department of Physics, University of Konstanz, D-78457 Konstanz, Germany}

\author{\firstname{Guido}~\surname{Burkard}}
 \email{guido.burkard@uni-konstanz.de}
\affiliation{Department of Physics, University of Konstanz, D-78457 Konstanz, Germany}

\begin{abstract}
    Dispersively coupled distant qubits in a shared cavity can become entangled through virtual photon exchange with energy-conserving phase evolution of their quantum states. This interaction can potentially be accelerated by operating on resonance, allowing for the exchange of real photons. In this theoretical study, we examine photon-mediated entanglement between two distant spins of electrons confined in double quantum dots formed in a Si/SiGe heterostructure. We calculate the dynamics of the combined system comprised of both spin qubits and the cavity, assuming that both spin qubits can be tuned into and out of resonance with the host cavity. We demonstrate that the exchange of real photons between the two spin qubits can result in rapid entanglement that is robust against decoherence. These results pave the way for the development of quantum gates on resonantly coupled distant semiconductor spin qubits. 
\end{abstract}
\maketitle

\section{Introduction}
A semiconductor quantum dot (QD) filled with an electron or hole represents a natural realization of a well-defined spin qubit (SQ)~\cite{Loss1998}, which has remained the subject of extensive studies for decades~\cite{Burkard2023}. 
Extending the system by a second QD to a double quantum dot (DQD) allows for electrical manipulation of charge~\cite{Hayashi2003, Petta2004, Dovzhenko2011, Cao2013, Kayatz2024} and spin~\cite{Cottet2010, Russ2018, Benito2019Flopping, Croot2020, Mutter2021} qubits, and controllable qubit-cavity coupling~\cite{Xuedong2012, Benito2017, Mi2017, Mi2018, Samkharadze2018, Dijkema2025}. 
The first experimental evidence of a measurable coherent interaction between an electron charge in a quantum dot and the quantized electromagnetic field of a superconducting resonator were achieved in the optical regime~\cite{Reithmaier2004, Yoshie2004} and paved the way for the engineering of the inter-qubit photon-mediated interaction~\cite{Borjans2020, Harvey2022}. 

A qubit with transition frequency $E_\sigma$ coupled to a superconducting microwave cavity mode $\omega$ with a coupling strength $g$ can operate in two distinct regimes of cavity quantum electrodynamics (QED). In the dispersive limit $|E_\sigma-\omega|\gg g$, the qubit is weakly coupled to the cavity field, and emission or absorption of photons are suppressed, while in the resonant regime $|E_\sigma-\omega|\ll g$, excitations can be exchanged between a qubit and the cavity mode. Operating in the dispersive regime at low temperatures, the number of photons in a cavity can be assumed to be zero~\cite{Benito2017}. In this case, the interaction between distant qubits is given by an effective Hamiltonian, which allows for the implementation of the $i\mathrm{SWAP}$~\cite{Hu2023, Kayatz2024} and CNOT~\cite{Warren2021, McMillan2023} two-qubit gates. Experiments have proved that the entanglement can be realized with fidelities up to 84\% in $21$~ns between qubits located at distance $250$~$\upmu$m~\cite{Dijkema2025}. A significant gate acceleration was predicted for qubits with frequencies tuned to the cavity photons energy~\cite{Woerkom2018, Marlon2021}. In this resonant regime, electrons exchange real photon and can become maximally entangled in only $\sim 10$~ns~\cite{Dijkema2025}. The main challenge the researchers face developing gates for charge qubits is the relatively high decoherence rate~\cite{Petersson2010}, which is caused by the high sensitivity of the electron wave function to charge noise. This circumstance motivates expanding the scope of cavity-based hybrid systems to include semiconductor spin qubits, which serve as a promising platform for programmable quantum processors~\cite{Zajac2018, Bradley2019, John2025}. To facilitate spin-photon interaction, the coupling between a polar vector (the electric field) and an axial pseudovector (the electron spin) is required. The natural realization of this coupling is provided by the spin-orbit interaction that contributes to the Hamiltonian with terms proportional to $\mathbf{k} \cdot \boldsymbol{\sigma}$, where $\mathbf{k}$ is electron's wave vector, and $\boldsymbol{\sigma}$ is the vector of Pauli spin operators. The standard electric dipole spin resonance (EDSR) technique employs the spin-orbit coupling~\cite{Golovach2006} and remains a widely used method in semiconductor spin-qubit experiments~\cite{Nadj-Perge2010, Schroer2011, Yu2023, Rimbach2025}. Since Si is not distinguished by a strong spin-orbit interaction strength, an artificial coupling between the DQD electric dipole and spin can be achieved via the introduction of a transverse non-uniform magnetic field with different directions in the two sites of a DQD, thereby inducing a spin-charge hybridization~\cite{Pioro2008}. Experiments have shown that the application of a magnetic field gradient to a single electron confined within a DQD leads to a coherent coupling between spin and a cavity photon~\cite{Mi2018, Samkharadze2018}, as predicted theoretically several years before~\cite{Benito2017}. Similar to charge qubits, $\sqrt{i\mathrm{SWAP}}$ and $i\mathrm{SWAP}$ gates were theoretically studied for two distant electron spins weakly coupled to the shared cavity~\cite{Benito2019}. Recent experiments have demonstrated a significant strengthening of the inter-spins coupling beyond the dispersive regime, when the photon energy matches the spin splitting of spin qubits (SQs)~\cite{Borjans2020}. A recent proposal for parametrically driven entanglement between spin qubits employs the sideband resonances~\cite{Srinivasa2024}. However, the theoretical picture of resonant photon-mediated spin-qubit coupling is still lacking. 

In this paper, we present a theoretical study of two spin qubits placed in a superconducting cavity tuned to the frequency of a single cavity mode. The transverse coupling of spin qubits to the cavity is taken into account within the first-order perturbation theory by a Schrieffer-Wolff transformation and by moving to the polariton basis. A rapid implementation of the $\ket{\uparrow \downarrow} \rightarrow \frac{1}{\sqrt{2}}(\ket{\uparrow \downarrow} - \ket{\downarrow \uparrow})$ transition is predicted for an appropriate sequence of voltage pulses applied to the electrodes to temporally detune the spin qubits from the cavity. Analyzing the free evolution of the system from a predetermined initial state $\ket{\uparrow \downarrow}$ reveals the existence of the most optimal degree of spin-charge hybridization, which corresponds to a rapid emergence of a maximally entangled state. Additional calculations demonstrate the minimal impact of relaxation and noise on the proposed two-spin qubit entanglement processes. 

The rest of the paper is organized as follows. In Sec.~\ref{sec:model} we present the model including the Hamiltonian in the dressed states basis and the required transformations. Sec.~\ref{sec:2step} describes the process of entangling two SQs from a given initial state using voltage pulses to turn the interaction between the SQs and the cavity on and off. Sec.~\ref{sec:combined_evolution} discusses the evolution of the three-partite system comprising two SQs and the cavity from the predetermined initial state and determines the conditions for maximizing entanglement between the qubits. Sec.~\ref{sec:conclusions} concludes the paper with a brief summary of the obtained results and open questions.  

\section{Model \label{sec:model}}
\begin{figure}
    \centering
    \includegraphics[width=1\linewidth]{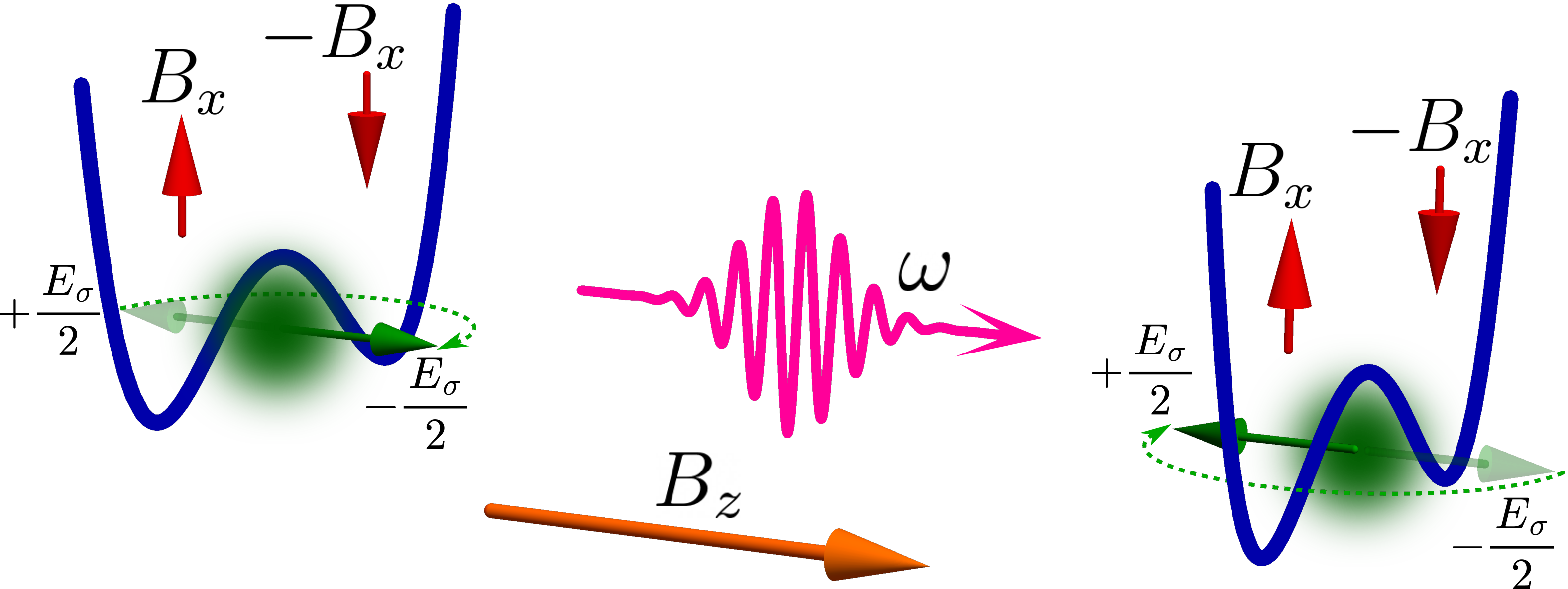}
    \caption{ Two electrons (green clouds) are confined within the DQDs in the shared cavity. A strong magnetic field $B_z$ lifts the spin degeneracy, while a relatively weak non-uniform field $\pm B_x$ hybridizes the spin and charge electronic states. A real photon with energy $\omega$ is transferred from one spin qubit to another and induces spin flips of both electrons whose energy levels are tuned to fulfill the resonance condition $E_\sigma = \omega$. }
    \label{fig:fig1}
\end{figure}
We consider two distant identical DQDs formed with metallic electrodes placed over the Si/SiGe heterostructure (Fig.~\ref{fig:fig1}). Each DQD is assumed to be symmetric and filled with one electron. While a uniform magnetic field $B_z$ causes a substantial Zeeman splitting, the spatially varying magnetic field provides relatively weak antiparallel fields $\pm B_x$ for the left and right sides of each DQD. The Hamiltonian of an electron in such a DQD can be expressed in the basis $\ket{L,+\frac{1}{2}}, \ket{L,-\frac{1}{2}}, \ket{R,+\frac{1}{2}}, \ket{R,-\frac{1}{2}}$, where $\ket{L}$ and $\ket{R}$ denote the states of an electron located in the left and right site of the DQD, and $\ket{\pm \frac{1}{2}}$ correspond to the electron spin orientation parallel ($-$) and antiparallel ($+$) to the field $B_z$~\cite{Burkard2020},
\begin{equation}
    \mathcal{H}_\mathrm{SQ} = \frac{B_z}{2} \widetilde{\sigma}_z + \frac{B_x}{2} \widetilde{\sigma}_x \widetilde{\tau}_z + t_c \widetilde{\tau}_x,
    \label{eq:SQ_H}
\end{equation}
where $t_c$ is the real-valued intradot tunneling strength, $B_x$ is the coupling coefficient between different spin states, and $\widetilde{\sigma}$ and $\widetilde{\tau}$ are the Pauli matrices acting in the $\ket{+\frac{1}{2}}$, $\ket{-\frac{1}{2}}$ and $\ket{L}$, $\ket{R}$ subspaces, respectively. When the spin qubit is placed into a microwave cavity with resonant frequency $\omega$, the linear interaction between the spin qubit and the electric displacement $D = E_0 (a + a^\dagger)$ is described by the Hamiltonian 
\begin{equation}
    \mathcal{H}_\mathrm{int} = E_0 e d (a + a^\dagger) \widetilde{\tau}_z,
\end{equation}
where $E_0$ is the one-photon electric field strength inside the DQD, $a^\dagger$ and $a$ are photon creation and annihilation operators, $e$ is elementary charge, and $d$ is the interdot spacing. Notably the populations of the $\frac{1}{\sqrt{2}} (\ket{L} \pm \ket{R})$ states oscillate with the vacuum Rabi frequency $g= E_0 e d$ when the DQD is brought into the interaction with a cavity with frequency $\omega=2t_c$. The resultant Hamiltonian $\sum_{k = 1,2} \left(\mathcal{H}_{\mathrm{SQ}_k} + \mathcal{H}_{\mathrm{int}_k}\right) + \omega a^\dagger a$ of the system comprising two spin qubits numbered with $k$ and a cavity is commonly transformed into the dressed-states basis, which diagonalizes the part $\sum_{k=1,2} \mathcal{H}_{\mathrm{SQ}_k}$, and where the interaction $\sum_{k=1,2} \mathcal{H}_{\mathrm{int}_k}$
is a small off-diagonal perturbation. In this basis, the system Hamiltonian takes the form~\cite{Benito2017}
\begin{equation}
\begin{gathered}
    \mathcal{H} = \omega a^\dagger a    + \sum_{k=1,2}\left[ \frac{E_\sigma}{2} \sigma_z^{(k)} + \frac{E_\tau}{2} \tau_z^{(k)}\right. \\
    \left. + (g_\sigma \sigma_x^{(k)} \tau_z^{(k)} - g_\tau \tau_x^{(k)})(a + a^\dagger) 
    \right],
\end{gathered}
\label{eq:H}
\end{equation}
where $2E_{\tau, \sigma} = \sqrt{(2 t_c + B_z)^2 + B_x^2} \pm \sqrt{(2 t_c - B_z)^2 + B_x^2}$, $g_\tau = g \cos{\phi}$, $g_\sigma = g \sin{\phi}$, and $\phi = \frac{\phi_+ + \phi_-}{2}$ with $\phi_\pm = \arctan{\displaystyle\frac{B_x}{2t_c \pm B_z}}$. To define the introduced $\sigma^{(k)}$ and $\tau^{(k)}$ matrices, we  describe their eigenstates, which are: $\ket{+ \uparrow}, \ket{+ \downarrow}, \ket{- \uparrow}$, and $\ket{- \downarrow}$. Here, the $(+)-$ signs denote the (anti)bonding orbitals $\ket{\pm} = \frac{1}{\sqrt{2}}(\ket{L} \pm \ket{R})$ with energies $\pm E_\tau/2$, while $\tau_z \ket{\pm} = \pm \ket{\pm}$ is the Pauli matrix for the orbital subspace. By analogy, $\sigma_z$ acts in the spin-like subspace $\sigma_z \ket{\uparrow (\downarrow)} = \pm \ket{\uparrow (\downarrow)}$. Explicitly,
\begin{eqnarray}
    \ket{-\downarrow} & \approx & \ket{-} \otimes \ket{-1/2}, \nonumber \\
        \ket{-\uparrow} & = & \cos{\frac{\phi}{2}} \ket{-}\otimes\ket{+1/2} +\sin{\frac{\phi}{2}} \ket{+}\otimes\ket{-1/2}, \nonumber \\
        \ket{+\downarrow} & = & -\sin{\frac{\phi}{2}} \ket{-}\otimes\ket{+1/2} +\cos{\frac{\phi}{2}} \ket{+}\otimes\ket{-1/2}, \nonumber \\
        \ket{+\uparrow} & \approx & \ket{+} \otimes \ket{+1/2},
\end{eqnarray}
where in the first and fourth equations we assumed that $\sqrt{(2t_c-B_z)^2 + B_x^2} / (2t_c+B_z) \ll 1$~\cite{Benito2017}. We denote the basis states of the Hamiltonian presented in Eq.~\eqref{eq:H} as $\ket{\pm \uparrow(\downarrow),\pm \uparrow(\downarrow)}_n$, where the orbital and spin states of both qubits are shown together with the number $n$ of photons in the cavity. Notwithstanding the fact that the $\ket{\uparrow}$ and $\ket{\downarrow}$ states do not exactly correspond to the electron spin projection, we refer to them as the spin-up and spin-down states that commonly encode logic states of the described qubits~\cite{Dijkema2025}. 

In the future framework, we focus on the resonant case $E_\sigma = \omega$, $E_\tau = \omega + \Delta$, which is the most promising for efficient photon-mediated spin control. This condition leads to a number of degenerate levels, such as $\ket{-\uparrow, -\uparrow}_0$, $\ket{-\downarrow, -\uparrow}_1$, $\ket{-\uparrow, -\downarrow}_1$, and $\ket{-\downarrow, -\downarrow}_2$. While in the dispersive regime one commonly performs a Schrieffer-Wolff (SW) transformation and moves to the empty cavity limit to derive the effective Hamiltonian for the inter-SQ interaction~\cite{Benito2019}, here, the photonic degree of freedom cannot be excluded. Indeed, when spin qubits are in resonance with the cavity, a considerable part of their energy will be transferred to the photonic form. Thus, the cavity field is a third essential component of the system. First, we perform a SW transformation for the Hamiltonian $\mathcal{H} = \mathcal{H}_0 + V_\mathrm{res} + V_\mathrm{disp}$, where $\mathcal{H}_0$ is the diagonal part of $\mathcal{H}$, while
\begin{equation}
    V_\mathrm{res} = g_\sigma \sum_{k=1,2} \left(a \sigma_+^{(k)} \tau_z^{(k)} + a^\dagger \sigma_-^{(k)} \tau_z^{(k)}\right)
\end{equation}
and
\begin{equation}
\begin{gathered}
    V_\mathrm{disp} = \sum_{k=1,2} \Big( g_\sigma (a \sigma_-^{(k)} \tau_z^{(k)} + a^\dagger \sigma_+^{(k)} \tau_z^{(k)}) \\
    - g_\tau \tau_x^{(k)} (a+a^\dagger) \Big)
\end{gathered}
\end{equation}
represent block-diagonal part and off-diagonal blocks of the qubit-cavity interaction Hamiltonian, respectively. Using the relation $V_\mathrm{disp} + [S, \mathcal{H}_0] = 0$ we find the SW generator~\cite{Rukhsan2020} 
\begin{eqnarray}
    S & = \displaystyle\sum_{k=1,2} \Big[ \dfrac{g_\tau}{\Delta}(-a \tau_+^{(k)} + a^\dagger \tau_-^{(k)}) \nonumber \\
    & -\dfrac{g_\tau}{2\omega + \Delta}(-a \tau_-^{(k)} + a^\dagger \tau_+^{(k)}) \\
  & + \dfrac{g_\sigma}{2\omega} (-a \sigma_-^{(k)} + a^\dagger \sigma_+^{(k)} ) \tau_z^{(k)} \Big] \nonumber
\end{eqnarray}
and expand the transformed Hamiltonian $\mathcal{H}' = e^S \mathcal{H} e^{-S}$ to the first order of $S$
\begin{equation}
    \mathcal{H}' \approx \mathcal{H}_0 + V_\mathrm{res} + \frac{1}{2}[S, 2 V_\mathrm{res} + V_\mathrm{disp}].
    \label{eq:res}
\end{equation}
Note that $\mathcal{H}$ contains degenerate blocks, which acquire additional off-diagonal elements after the SW transformation. For example, there is no direct transition between states $\ket{- \uparrow, - \uparrow}_0$ and $\ket{- \uparrow, - \uparrow}_0$ in $\mathcal{H}$, but they both are coupled to the transitional state $\ket{- \uparrow, + \uparrow}_1$. The SW transformation excludes the $\ket{- \uparrow, + \uparrow}_1$ state, but introduces an off-diagonal element coupling the degenerate states $\ket{- \uparrow, - \uparrow}_0$ and $\ket{- \uparrow, - \uparrow}_0$. Importantly, the levels bound by the potential $V_\mathrm{res}$ in $\mathcal{H}$ are no longer degenerate in $\mathcal{H}'$, because of the small SW corrections $\sim \frac{g^2}{\Delta}$. However, when $g_\sigma \gg \frac{g_\tau^2}{\Delta}$, these corrections can be neglected. Therefore, to diagonalize $\mathcal{H}'$ we introduce a relatively simple rotation matrix $R$, the application of which entangles spin and photonic states and transforms the Hamiltonian to the polariton basis: $\widetilde{\mathcal{H}} = R^T\mathcal{H}'R$. As a result, the initial Hamiltonian from Eq.~\eqref{eq:H} is transformed by the unitary $U = e^{-S} R$ to the first order of $S$. For simplicity, we assume that the Hilbert space is rotated by $R$ only, because $U \approx (\mathbb{1} - S)R$, and $S \ll \mathbb{1}$. Thus, transitional states are excluded from our consideration, but their influence on the energies of other states is accounted for by Eq.~\eqref{eq:res}. 

To quantify the influence of decoherence on the open system state, we solve the master equation with the Lindbladian in the Markovian limit, 
\begin{eqnarray}
    \dot{\rho} &=& -i[\mathcal{H}, \rho] +  \kappa \mathcal{L}[a] \rho  \label{eq:master} \\
    & & + \sum_{k=1,2} \left(\frac{\gamma_\phi}{4} \sin^2{\phi} \mathcal{L}[\sigma_z^{(k)}] \rho + \frac{\gamma}{2} \sin^2{\phi} \mathcal{L}[\sigma_-^{(k)}] \rho \right), \nonumber
\end{eqnarray}
for the density matrix $\rho$, where $\gamma_\phi$ and $\kappa$ are the charge-noise-induced dephasing rate  of the spin qubit and cavity damping coefficient, $\gamma$ characterizes the phonon-induced charge relaxation, and $\mathcal{L}[A] \rho = 2 A \rho A^\dagger - \rho A^\dagger A - A^\dagger A \rho$ stands for the dissipator given by the quantum jump operator $A$. 

\section{Two-step entangling \label{sec:2step}}
\begin{figure}
    \centering
    \includegraphics[width=1.\linewidth]{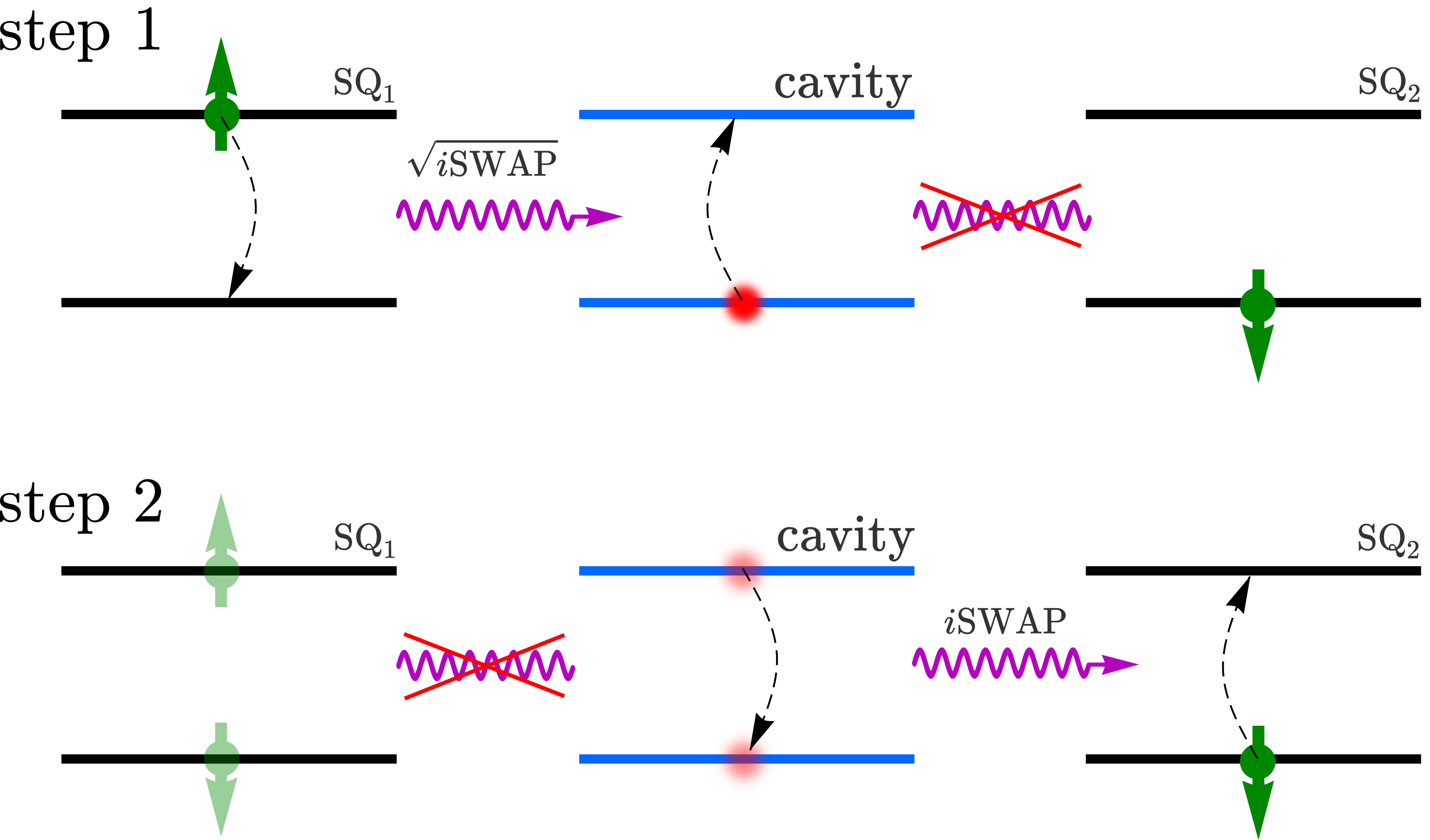}
    \caption{ Two-step entanglement scheme for the initial state $\ket{\Psi_0} = \ket{\text{SQ}_1 \text{SQ}_2}_0 = \ket{\uparrow \downarrow}_0$ with empty cavity, the first spin qubit (SQ$_1$) prepared in the excited $\ket{\uparrow}$ state and the second spin qubit (SQ$_2$) in the ground $\ket{\downarrow}$ state. In the first step, the ground-state spin qubit (SQ$_2$) is decoupled, whereas the $\sqrt{i\mathrm{SWAP}}$ operation is performed between the initially excited spin qubit (SQ$_1$) and the cavity. The second step is the subsequent $i\mathrm{SWAP}$ between SQ$_2$ and the cavity, leaving the cavity empty.}
    \label{fig:scheme}
\end{figure}
In the dispersive regime, the effective approximate Hamiltonian of the form $\propto \frac{g_\sigma^2}{E_\sigma-\omega}(\sigma_+^{(1)}\sigma_-^{(2)} + \sigma_-^{(1)}\sigma_+^{(2)})$ results in $i\mathrm{SWAP}$ gate between the spin qubits within a time $\sim \pi \frac{E_\sigma-\omega}{2 g_\sigma^2}$. Although the instantaneous energy of the electromagnetic field is negligible, virtual microwave photons transfer energy from one qubit to the other in small amounts, entangling them. After a relatively long time, when sufficient energy exchange is performed, the cavity field does generate a quantum gate operation for the spin qubits. In the resonant case studied here, $i\mathrm{SWAP}$ gates with a rate $\sim g_\sigma$ run simultaneously between both spin qubits and the cavity, which is filled with real photons that carry large amounts of energy. Thus, the energy flux between the qubits is the maximum possible, but fast entanglement is not guaranteed because the cavity becomes empty at certain moments in time only. By contrast, simple vacuum Rabi oscillations are expected for the interaction between single qubits and the cavity. In order to utilize the channels between different spin qubits and resonator separately, we propose switching the qubit-cavity interaction on and off. This can be realized via an instant voltage-induced increase of the DQD detuning~\cite{Harvey2022}. Consider the initial state $\ket{\Psi_0} = \ket{-\uparrow,-\downarrow}_0$ with the second ($\downarrow$) qubit detuned from the cavity. In this case, the state of the system evolves to
\begin{equation}
    \ket{\Psi(\tau_1)} = \cos{(g_\sigma \tau_1)} \ket{\uparrow \downarrow}_0 + i \sin{(g_\sigma \tau_1)} \ket{\downarrow \downarrow}_1
\end{equation}
after time $\tau_1$, where we omitted the orbital states $\ket{-,-}$. At this moment, SQ$_1$ is deactivated, while SQ$_2$ is brought to the resonance with the cavity for the time period $\tau_2$. As a result, after the time $\tau_1+\tau_2$ the system appears in the state 
\begin{eqnarray}
     \ket{\Psi(\tau_1+\tau_2)} &=& \cos{(g_\sigma \tau_1)} \ket{\uparrow \downarrow}_0 \nonumber \\
    & &- \sin{(g_\sigma \tau_1)} \sin{(g_\sigma \tau_2)} \ket{\downarrow \uparrow}_0 \label{eq:state_tau1-tau2} \\
    & & + i \sin{(g_\sigma \tau_1)}  \cos{(g_\sigma \tau_2)}  \ket{\downarrow \downarrow}_1.\nonumber
\end{eqnarray}
Remarkably, at $\tau_2 = 2 \tau_1 = \frac{\pi}{2 g_\sigma}$, the qubits arrive at the maximally entangled Bell state $\frac{1}{\sqrt{2}} (\ket{\uparrow \downarrow}_0 - \ket{\downarrow \uparrow}_0)$. As illustrated in Fig.~\ref{fig:scheme}, the route from the initial state $\ket{\Psi_0}=\ket{\uparrow \downarrow}_0$ to $\ket{\Psi(\tau_1+\tau_2)} = \frac{1}{\sqrt{2}} (\ket{\uparrow \downarrow}_0 - \ket{\downarrow \uparrow}_0)$ consists of two separate gates: $\sqrt{i \mathrm{SWAP}}$ between SQ$_1$ and cavity and $i \mathrm{SWAP}$ between cavity and SQ$_2$. These two gates implemented sequentially lead to the $\ket{\uparrow \downarrow} \rightarrow \frac{1}{2}(\ket{\uparrow \downarrow} - \ket{\downarrow \uparrow})$ rotation. 
\begin{figure}
    \centering
    \includegraphics[width=0.8\linewidth]{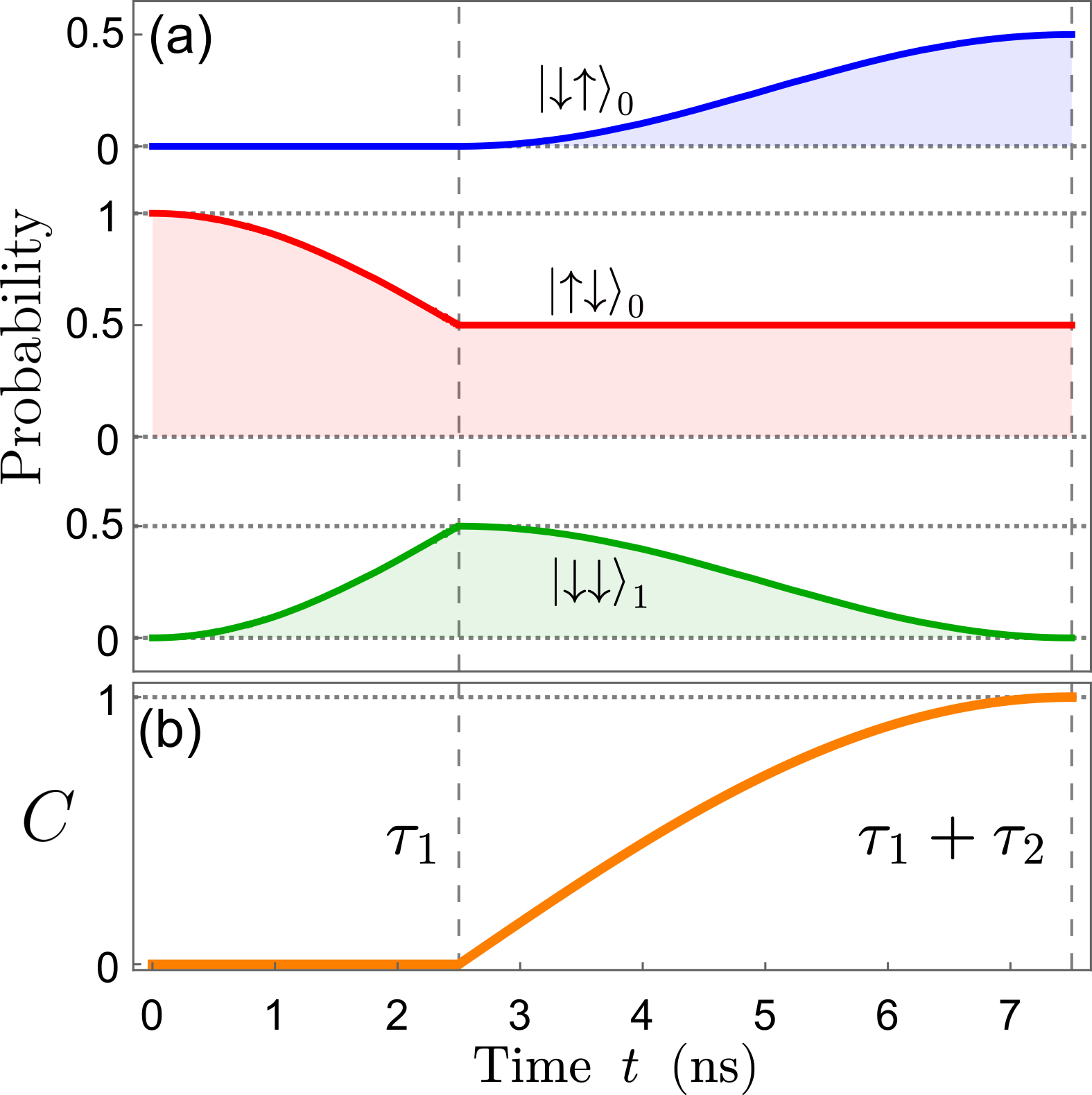}
    \caption{The process of entangling the two spin qubits. The panel (a) depicts the probabilities of the two-qubit system to occupy different basis states, while the panel (b) shows the temporal evolution of the entanglement measured via the concurrence $C$. The parameters used in the computation are $\omega / 2\pi = 10$~GHz, $\Delta / 2\pi = 2$~GHz, $g / 2\pi = 100$~MHz, $\phi = \pi/6$. }
    \label{fig:fast-entangled}
\end{figure}
We quantify the amount of entanglement using the concurrence $C = \mathrm{max}(0, \lambda_1 - \lambda_2 - \lambda_3 - \lambda_4)$~\cite{Wootters1998} applied to the reduced density matrix $\rho_r$ of the two spin qubits. Here, $\lambda_i$ are the eigenvalues of the matrix $\sqrt{\sqrt{\rho_r} \widetilde{\rho}_r \sqrt{\rho_r}}$ with $\widetilde{\rho}_r = (\sigma_y \otimes \sigma_y)\rho_r^*(\sigma_y \otimes \sigma_y)$.
In Fig.~\ref{fig:fast-entangled} we present the calculated time evolution of the quantum state as well as its concurrence
\begin{equation}
    C = \begin{cases}
        0, \; \text{if} \; t < \tau_1,\\
        |\sin{(2 g_\sigma \tau_1)} \sin{[g_\sigma (t - \tau_1) ]}|, \; \text{if} \; t \geq \tau_1
    \end{cases},
\end{equation}
which is derived from Eq.~\eqref{eq:state_tau1-tau2}. We briefly note that there is another variant of two-step entanglement: do $i \mathrm{SWAP}$ between the first qubit and cavity, and then let both qubits fully absorb the emitted photon. However, since the main source of decoherence is the field leakage from the cavity, the resultant entanglement would be much less, because photon loss would drive the system from the state $\ket{\downarrow \downarrow}_1$ to the state $\ket{\downarrow \downarrow}_0$. 

Next, we estimate the influence of relaxation and dephasing on the entanglement procedure suggested above. To this end, we solve Eq.~\eqref{eq:master} both analytically and numerically. Following the analogy with pure-state calculations, we restrict ourselves to the subspace with up to one excitation, spanned by the basis states $\ket{\downarrow \uparrow}_0, \ket{\uparrow \downarrow}_0, \ket{\downarrow \downarrow}_0, \ket{\downarrow \downarrow}_1$ and use of the simplified Hamiltonian (up to a multiple of the unity operator),
\begin{eqnarray}
    \mathcal{H} &=& - \omega \ket{\downarrow \downarrow}_0 \prescript{}{0}{\bra{\downarrow \downarrow}} \\
    & & + g_\sigma^{(1)}(t) \ket{\uparrow \downarrow}_0 \prescript{}{1}{\bra{\downarrow \downarrow}} + g_\sigma^{(2)}(t) \ket{\downarrow \uparrow}_0 \prescript{}{1}{\bra{\downarrow \downarrow}} + \mathrm{h.c.} \nonumber,
\end{eqnarray}
where $g_\sigma^{(1)} = g_\sigma$ and $g_\sigma^{(2)} = 0$ for $t < \tau_1$, and $g_\sigma^{(1)} = 0$ and $g_\sigma^{(2)} = g_\sigma$ for $t > \tau_1$. Restricting it to the four states only is justified by choosing values of $\tau_1$ and $\tau_2$ that are much smaller than $\frac{2\pi}{g^2/\Delta} \sim 50-100$~ns. Furthermore, we neglect the terms proportional to $\gamma$, $\gamma_\phi$, and $\kappa$ in all $\dot{\rho}$ elements except for $\prescript{}{0}{\bra{\uparrow \downarrow}} \dot{\rho} \ket{\downarrow \downarrow}_1$ and $\prescript{}{1}{\bra{\downarrow \downarrow}} \dot{\rho} \ket{\uparrow \downarrow}_0$ that determine the degree of the state purity. Note that this truncation implies that the master equation is not trace-preserving, but at relatively small times the reduction of the trace is negligible. After the time $\tau_1$ of interaction between the first spin qubit and the cavity, the density matrix becomes
    \begin{eqnarray}
        \rho(\tau_1) = \frac{1}{2}\left( \ket{\uparrow \downarrow}_0 \prescript{}{0}{\bra{\uparrow \downarrow}} + \ket{\downarrow \downarrow}_1 \prescript{}{1}{\bra{\downarrow \downarrow}} \right) \nonumber \\
        \frac{e^{-\lambda \tau_1}}{2} \Big[ (1+\cos{2 g_\sigma} \tau_1) \ket{\uparrow \downarrow}_0 \prescript{}{0}{\bra{\uparrow \downarrow}} \nonumber \\
        + (1-\cos{2 g_\sigma} \tau_1) \ket{\downarrow \downarrow}_1 \prescript{}{1}{\bra{\downarrow \downarrow}} \Big] \nonumber \\
        +\frac{i}{2} e^{-\lambda \tau_1} \sin{2 g_\sigma \tau_1} \ket{\uparrow \downarrow}_0 \prescript{}{1}{\bra{\downarrow \downarrow}} + \text{h.c.}, \label{eq:rho-tau1}
    \end{eqnarray}
where $\lambda = \frac{1}{4}(2\kappa + 2 \gamma_\phi \sin^2{\phi} + \gamma \sin^2{\phi})$ and we assumed that $\sqrt{g_\sigma^2-\frac{1}{16}(\kappa + \gamma_\phi \sin^2{\phi}+\frac{\gamma}{2} \sin^2{\phi})^2} \approx g_\sigma$. 
The probability of the $\ket{\downarrow \downarrow}_0$ state is $\prescript{}{0}{\bra{\downarrow \downarrow}} \rho(\tau_1) \ket{\downarrow \downarrow}_0 \approx (\kappa + \frac{\gamma}{2} \sin^2{\phi}) \tau_1 + \frac{\gamma \sin^2{\phi}-2\kappa}{4 g_\sigma} e^{-\lambda_1 \tau_1} \sin{2 g_\sigma \tau_1} \ll 1$. Eq.~\eqref{eq:rho-tau1} is then used as the initial condition for the master equation describing the evolution of the three-partite system with first qubit deactivated and second one interacting with the cavity. Note that, even when detuned, the first spin qubit still experiences charge noise and relaxation at a reduced rate. However, we use the same dissipator as for the spin qubit with zero detuning, which gives us a lower estimate of the resultant entanglement. After the time $\tau_2$ of the energy exchange between the second qubit and the cavity the system appears in the state with the nonzero independent $\rho$ elements listed in~\ref{sec:appendix} (we also omitted terms proportional to $\gamma$, $\gamma_\phi$, and $\kappa$ in $\prescript{}{0}{\bra{\downarrow \uparrow}} \dot{\rho} \ket{\downarrow \uparrow}_0$, $\prescript{}{1}{\bra{\downarrow \downarrow}} \dot{\rho} \ket{\downarrow \downarrow}_1$, $\prescript{}{0}{\bra{\downarrow \downarrow}} \dot{\rho} \ket{\downarrow \uparrow}_0$, $\prescript{}{0}{\bra{\downarrow \downarrow}} \dot{\rho} \ket{\uparrow \downarrow}_0$, $\prescript{}{0}{\bra{\downarrow \downarrow}} \dot{\rho} \ket{\downarrow \downarrow}_1$, and h.c. here). The probability $\prescript{}{0}{\bra{\downarrow \downarrow}}\rho \ket{\downarrow \downarrow}_0$ of the lowest energy $\ket{\downarrow \downarrow}_0$ state is found negligibly small at short $\tau_1$ and $\tau_2$ time intervals due to weak decoherence. Next, we investigate the state given by the Eq.~\eqref{eq:rho-tau1+tau2} via the calculation of the concurrence $C$ as a function of $\tau_1$ and $\tau_2$. Tracing the cavity states out, we derive the reduced density matrix $\rho_r$ of the two spin qubits and find the concurrence 
\begin{figure}[b!]
    \centering
    \includegraphics[width=1.\linewidth]{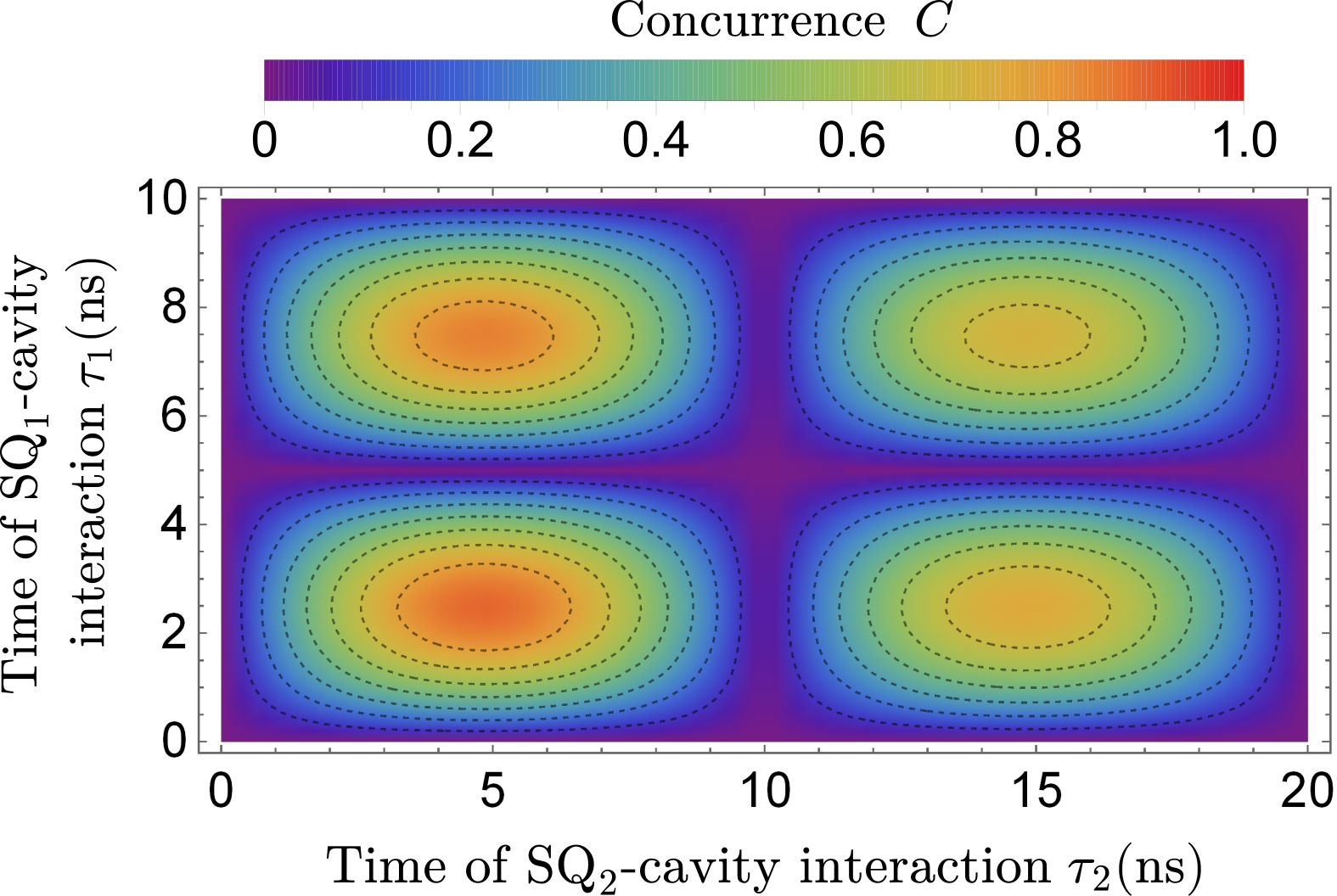}
    \caption{The concurrence from Eq.~\eqref{eq:Concurrence} as a function of two time intervals, during which SQ$_{1,2}$ are activated/deactivated. The maximum concurrence $C_\mathrm{max} = 0.9$ is found at $\tau_1 = 2.48$~ns and $\tau_2 = 4.82$~ns. The parameters used in the computation are $\omega / 2\pi = 10$~GHz, $\Delta / 2\pi = 2$~GHz, $g / 2\pi = 100$~MHz, $\phi = \pi/6$, $\kappa / 2 \pi = 2 $~MHz, $\gamma / 2 \pi = \gamma_\phi / 2 \pi = 3$~MHz. }
    \label{fig:C-t1-t2}
\end{figure}
\begin{equation}
    C = \big| \sin{(2 g_\sigma \tau_1)} \sin{(g_\sigma \tau_2)} \big| e^{-(\lambda \tau_1 + \Lambda \tau_2)}. 
    \label{eq:Concurrence}
\end{equation}
In Fig.~\ref{fig:C-t1-t2} the concurrence from Eq.~\eqref{eq:Concurrence} is plotted at representative numerical parameters values. The colour map contains a number of local maxima, which correspond to the $\sqrt{i \mathrm{SWAP}}$ gate between first qubit and the cavity followed by an integer number of $i \mathrm{SWAP}$ gates between first qubit and the cavity and odd number of $i \mathrm{SWAP}$ gates between the cavity and second qubit. The redundant $i\mathrm{SWAP}$ gates result in longer times $\tau_1$ or $\tau_2$ and the exponential decrease of the entanglement. Importantly, by solving the master equation numerically for the Hamiltonian from Eq.~\eqref{eq:H} with maximally possible photon number $n_\mathrm{max}=2$, we derive almost the same concurrence values. The deviation from the Eq.~\eqref{eq:Concurrence} is maximal (around 6\%) at $\tau_1 = 7.4$~ns and $\tau_2 = 4.8$~ns, and is less than 0.01\% near the global maximum. From Eq.~\eqref{eq:Concurrence} we find the time intervals corresponding to the maximal entanglement
\begin{equation}
    \begin{gathered}
        \tau_1  \approx \frac{\pi}{4 g_\sigma} - \frac{\lambda}{2 g_\sigma^2},\quad 
        \tau_2  \approx \frac{\pi}{2g_\sigma} - \frac{\Lambda}{g_\sigma^2}.
    \end{gathered}
\end{equation}
Decoherence causes a negative shift of both time intervals needed for maximal concurrence, which is given by
\begin{equation}
    C_\mathrm{max} \approx 1 - \frac{\pi}{4 g_\sigma}(\lambda+2 \Lambda).
\end{equation}
Interestingly, at weak spin-charge hybridization ($\phi \ll 1$) $\lambda \approx \Lambda \approx \kappa / 2$. Here, the cavity damping does not break the ratio $\tau_1 /\tau_2 = 2$ obtained at the maximal entanglement point. This feature can be explained by the fact that in both stages illustrated in Fig.~\ref{fig:fast-entangled} the average cavity relaxation rate is the same: $\int_0^{\tau_1} dt \prescript{}{1}{\bra{\downarrow \downarrow}} \dot{\rho} \ket{\downarrow \downarrow}_1 = \int_{\tau_1}^{\tau_2} dt \prescript{}{1}{\bra{\downarrow \downarrow}} \dot{\rho} \ket{\downarrow \downarrow}_1$. This conclusion is supported by comparing the probabilities of the state $\ket{\downarrow \downarrow}_0$ at time moments $\tau_1$ and $\tau_1+\tau_2$. 

\section{Spin qubits-cavity dynamics \label{sec:combined_evolution}}
\begin{figure}
    \centering
    \includegraphics[width=0.9\linewidth]{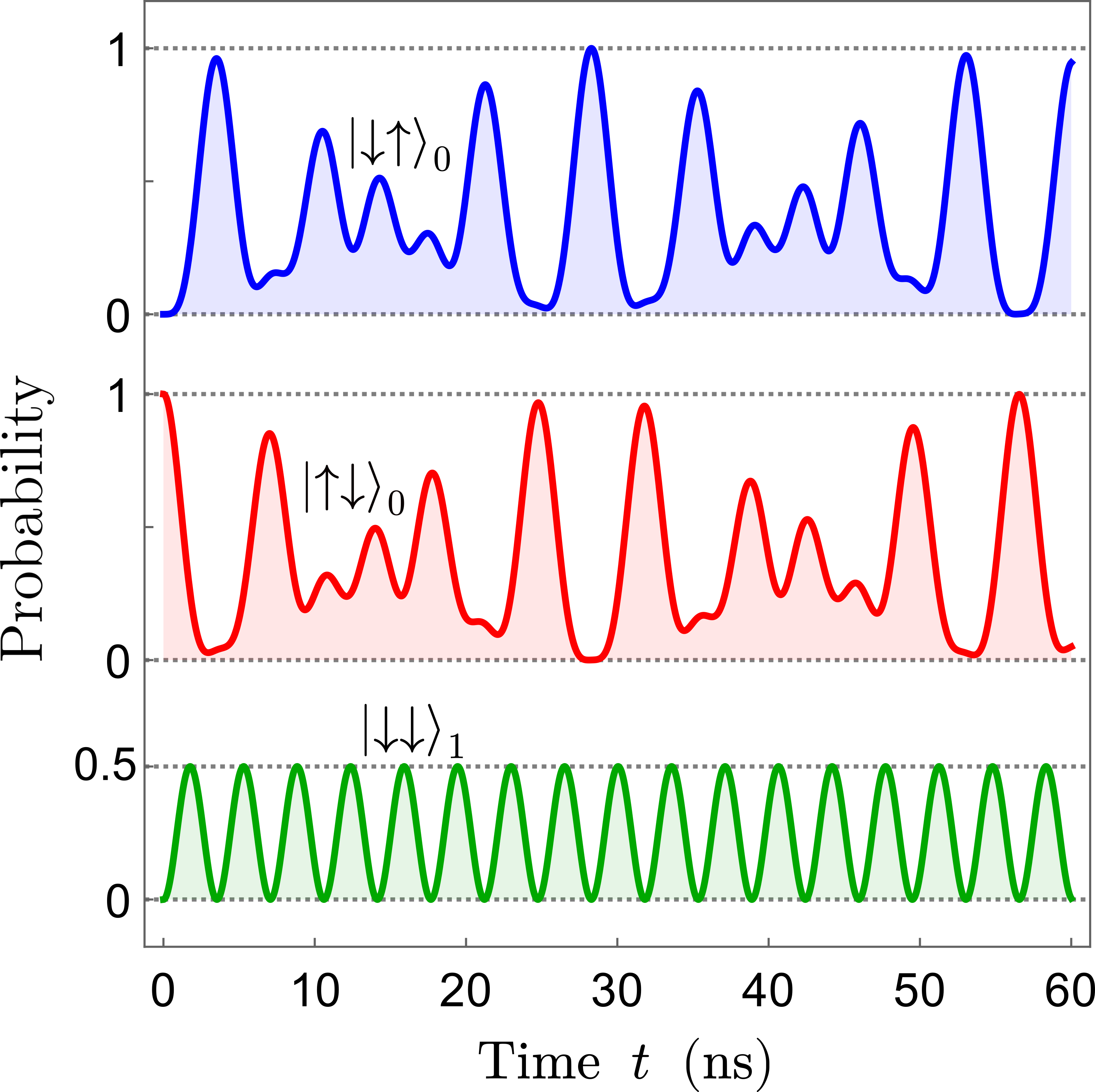}
    \caption{The temporal dependences of state probabilities for the initial condition $\ket{\Psi_0} = \ket{\uparrow \downarrow}_0$. The spin-charge angle is $\phi = \pi / 6$. }
    \label{fig:SWAP-P}
\end{figure}

\begin{figure}[t]
    \centering
    \includegraphics[width=0.85\linewidth]{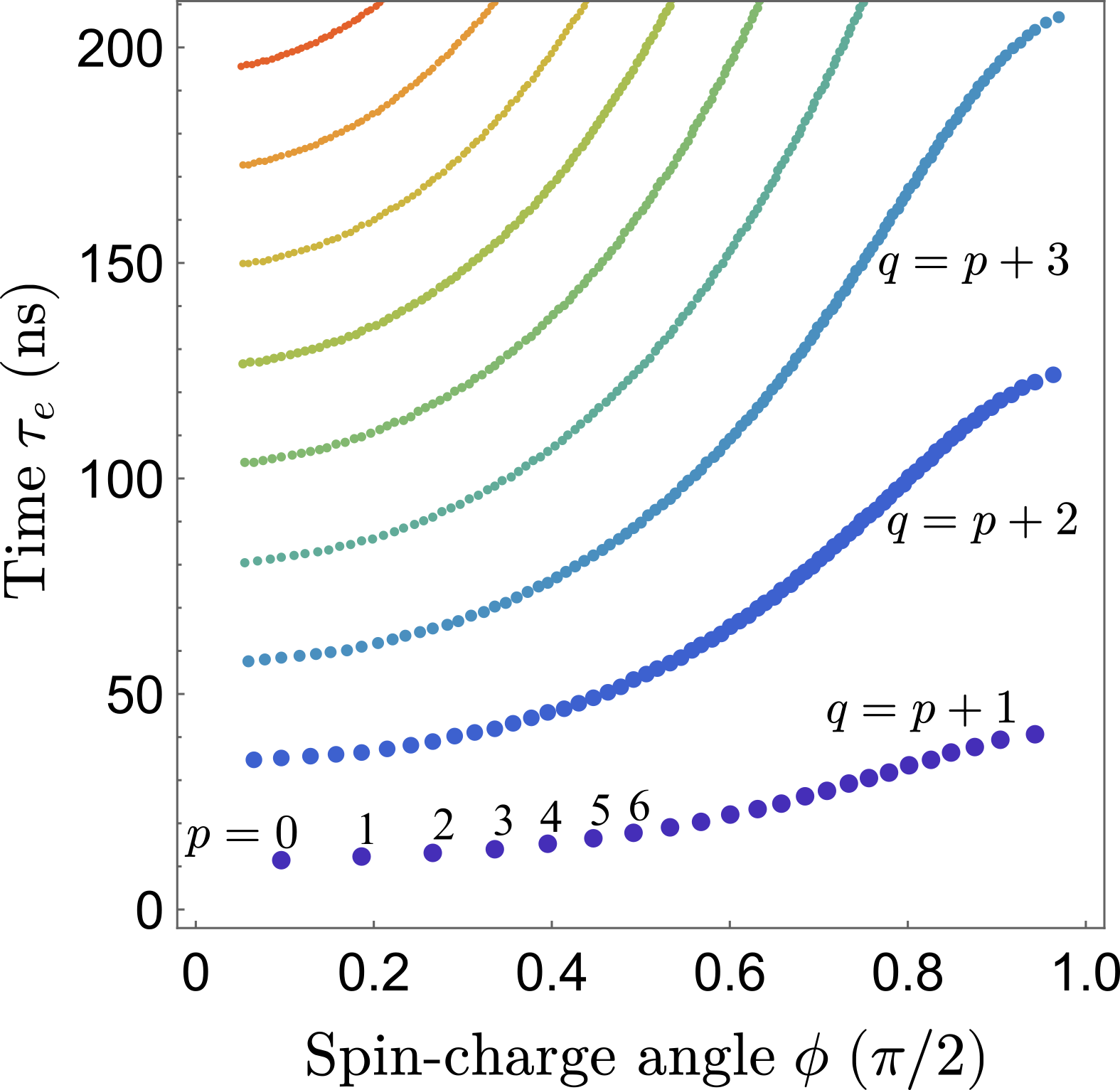}
    \caption{The points in the two-dimensional parameter space spanned by the spin-charge mixing angle $\phi$ and the time $\tau_e$ corresponding to the maximally entangled final state $\ket{\Psi(\tau_e)}$. The data is divided into several subsets with different $p-q$.}
    \label{fig:PQsets}
\end{figure}

In this section, we study the tripartite system evolution from the initial state $\ket{\Psi_0} = \ket{\uparrow\downarrow}_0$. The pure state of the combined three-partite system is found from the Hamiltonian $\widetilde{\mathcal{H}}$
\begin{eqnarray}
    \ket{\Psi} & = & \frac{1}{4}\left(e^{i(\varepsilon_1 + \sqrt{2} g_\sigma)t} + e^{i(\varepsilon_1 - \sqrt{2} g_\sigma)t} \right) ( \ket{\uparrow\downarrow}_0 + \ket{\downarrow\uparrow}_0 ) \nonumber \\
    & + & \frac{1}{2} e^{i \varepsilon_2 t} ( \ket{\uparrow\downarrow}_0 - \ket{ \downarrow\uparrow}_0 ) \label{eq:SWAP-solution} \\
    & + & \frac{1}{2\sqrt{2}} \left(e^{i(\varepsilon_1 + \sqrt{2} g_\sigma)t} - e^{i(\varepsilon_1 - \sqrt{2} g_\sigma)t} \right) \ket{\downarrow\downarrow}_1, \nonumber
\end{eqnarray}
where $\varepsilon_1 = \omega + \Delta + \frac{3 g_\sigma^2}{2 \omega} + \frac{4 g_\tau^2}{\Delta + 2 \omega}+\frac{2 \omega g_\tau^2}{2 \omega \Delta + \Delta^2}$, and $\varepsilon_2 = \omega+ \Delta +\frac{2 g_\tau^2}{\Delta + 2 \omega}$. In Eq.~\eqref{eq:SWAP-solution} every step of the Hamiltonian transformation presented in Sec.~\ref{sec:model} has left its imprint. In particular, the term $\frac{3g_\sigma^2}{2 \omega}$ comes from the SW transformation for the counter-rotating terms. By contrast, the correction $\frac{2\omega g_\tau^2}{2\omega \Delta +\Delta^2}$ reflects the influence of the rotating terms. Meanwhile, the splitting $\sqrt{2} g_\sigma$ corresponds to the spin-photon entangled eigenstates formed by the rotation $R$. In the following, we use the parameter values $\omega / 2\pi = 10$~GHz, $\Delta / 2\pi=2$~GHz, and $g / 2\pi = 200$~MHz~\cite{Dijkema2025}, unless stated otherwise. The correspondent probabilities of the basis states are presented in Fig.~\ref{fig:SWAP-P}, where after the transfer of energy about $\omega$ from the first qubit to the second, their spin states are almost interchanged.  
However, the derived state rotation does not exactly reproduce an $i$SWAP gate for the initial $\ket{\uparrow \downarrow}_0$ state. We attribute this imperfectness to the competition between different channels of energy transfer between SQs. Indeed, if there were only three possible states: $\ket{\uparrow \downarrow}_0$, $\ket{\downarrow \uparrow}_0$, and $\ket{\downarrow \downarrow}_1$, with two equivalent channels $\ket{\uparrow \downarrow}_0 \longleftrightarrow \ket{\downarrow \downarrow}_1$ and $\ket{\downarrow \uparrow}_0 \longleftrightarrow \ket{\downarrow \downarrow}_1$, the iSWAP gate would be perfect. However, the introduction of the additional degrees of freedom reflects itself in the extra energy channels $\ket{\uparrow \downarrow}_0 \longleftrightarrow \ket{\downarrow \uparrow}_0$, that compete with the main route $\ket{\uparrow \downarrow}_0 \longleftrightarrow \ket{\downarrow \downarrow}_1 \longleftrightarrow \ket{\downarrow \uparrow}_0$ of the energy flow. As a result, when the transitional state $\ket{\downarrow \downarrow}_1$ becomes empty, both states $\ket{\uparrow \downarrow}_0$ and $\ket{\downarrow \uparrow}_0$ are still occupied. In fact, after each period $\tau_\mathrm{SWAP} = \frac{\pi}{\sqrt{2} g_\sigma}$ of the $\ket{\downarrow \downarrow}_1$ state occupancy oscillation an $(i\mathrm{SWAP})^\alpha$ gate is performed. Within our approximation we find
\begin{equation}
    \alpha(\phi) = 1 - \frac{3 g_\sigma}{2 \sqrt{2} \omega } - \frac{\sqrt{2}(\omega+\Delta) g_\tau^2}{\Delta (2\omega+\Delta) g_\sigma}.
    \label{eq:alpha}
\end{equation}
Note that in Eq.~\eqref{eq:alpha} $\alpha$ is written as a function of $\phi$ at fixed parameters $E_\sigma = \omega$, $\Delta$, and $g$. This approach is justified by the ability to realize the introduced $\alpha(\phi)$ function by adjusting the parameters $B_z$, $B_x$, and $t_c$ to keep $E_\sigma = \omega$, $\Delta$, and $g$ constant.
Importantly, if the spin qubits undergo $q=1,2,3,...$ sequential $(i\mathrm{SWAP})^\alpha$ gates then the resultant $(i\mathrm{SWAP})^{q \alpha}$ operation is equivalent to $(i\mathrm{SWAP})^{p+1/2}$ with $p=0,1,2,...$. This transforms the initial $\ket{\uparrow\downarrow}_0$ state to the maximally entangled $\frac{1}{\sqrt{2}}(\ket{\uparrow \downarrow}_0 + e^{i (p+1/2) \pi} \ket{\downarrow \uparrow}_0)$ state.
Note that for different initial states, a maximally entangled state could also be obtained for integer $q\alpha$~\cite{Picard2025}. To obtain this in time $\tau_e = q \tau_\mathrm{SWAP}$, the condition $q \alpha(\phi) = p + \frac{1}{2}$ must be satisfied. After solving for all appropriate $\alpha(\phi)$ values, one can divide the derived set $\tau_e(\phi)$ into subsets with $q-p = 1,2,3, ...$, which are illustrated in Fig.~\ref{fig:PQsets}. 
Notably, the discrete functions $\tau_e(\phi, q-p)$ increase monotonically with respect to both arguments, which seems counter-intuitive, because $\phi$ determines the spin-photon coupling strength and its increase allows for faster gates. We find that the most optimal point $(p, q) = (0, 1)$ corresponds to the spin-charge angle $\phi \approx \frac{2\sqrt{2} g (\omega + \Delta)}{(2\omega + \Delta)\Delta} \sim \frac{g}{\Delta}$, which is surprisingly small. Meanwhile, the gate time is given by $\tau_e \approx \frac{\pi \Delta(2\omega + \Delta)}{4(\omega+\Delta)g^2}$ at this point. The $\tau_e \propto \Delta g^{-2}$ proportionality results in a drastic increase of $\tau_e$ from 12~ns at $g / 2\pi = 200$~MHz to 46~ns at $g / 2\pi = 100$~MHz. 

\begin{figure}
    \centering
    \includegraphics[width=0.8\linewidth]{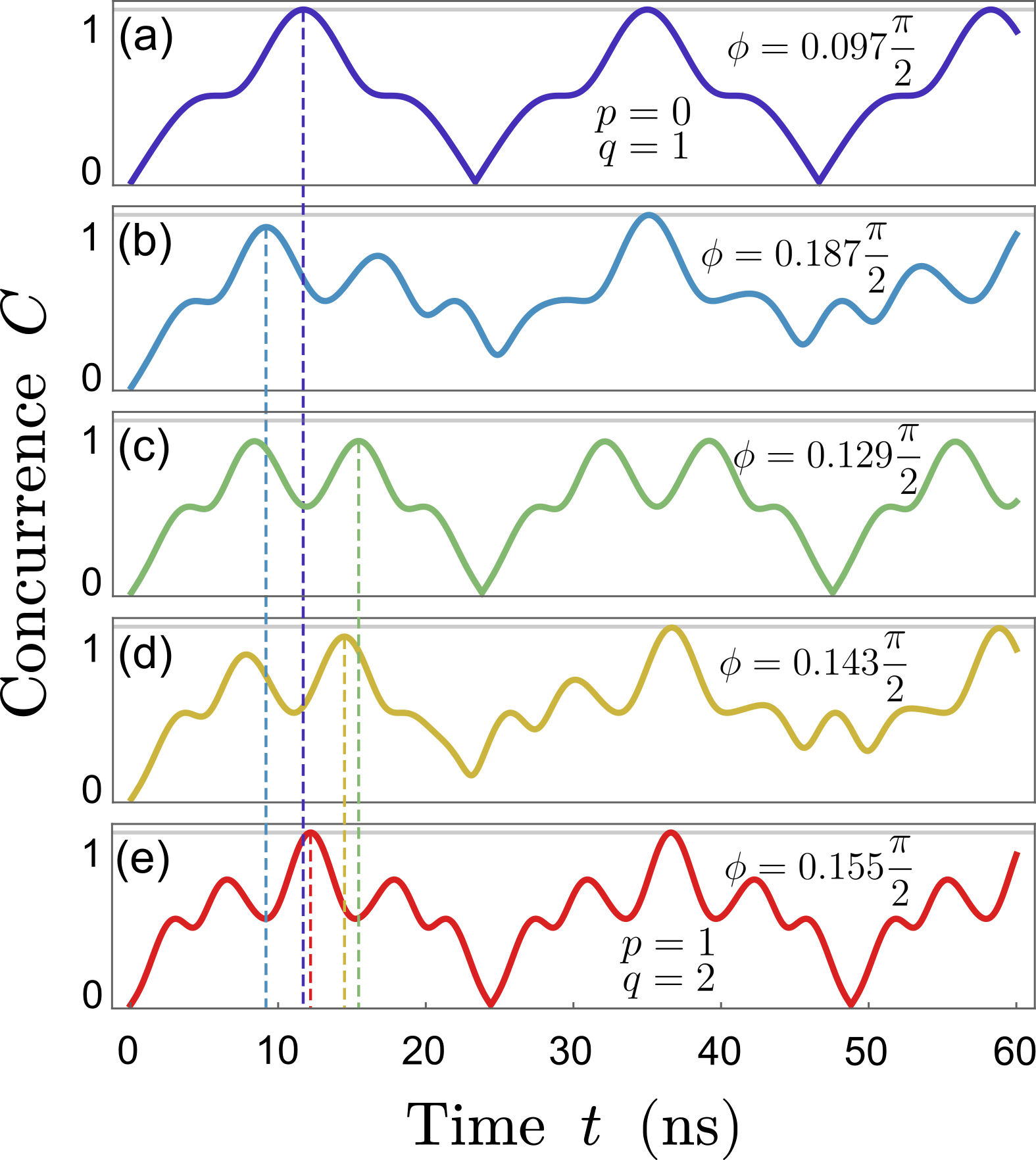}
    \caption{The concurrence $C$ plotted against time for different spin-charge angle values $\phi$. The dashed lines indicate the times at which the gates are completed.
    These are defined as the first local maximum within the series of local maxima of $C(t)$. (a) and (e) panels correspond to the first two points from the $q = p + 1$ branch in Fig.~\ref{fig:PQsets} ($\phi_{p=0,q=1}$, $\phi_{p=1,q=2}$), where $C_\mathrm{max} = 1$. The panels (b, c, d) illustrate the entanglement dynamics for $\phi_{p=0,q=1} < \phi < \phi_{p=0,q=2}$, where $C_\mathrm{max} < 1$. }
    \label{fig:C-t}
\end{figure}

Now, we focus on the $q = p + 1$ subset (the lowest branch in Fig.~\ref{fig:PQsets}), because it provides the minimal time $\tau_e$ for the entanglement between the SQs and is supposed to be the most robust against decoherence. To quantify the degree of entanglement in-between the discrete points we calculate the concurrence $C$ of the SQs given by the reduced density matrix is in the basis of the spin states $ \ket{\uparrow \uparrow}, \ket{\uparrow \downarrow}, \ket{\downarrow \uparrow}, \ket{\downarrow \downarrow}$. The expression follows immediately from Eq.~\eqref{eq:SWAP-solution},
\begin{equation}
\begin{gathered}
    C(t) = \frac{1}{4} \bigg| -1 + \cos{(2 \sqrt{2} g_\sigma t)}  \\
    + 2i \sin{(\sqrt{2} g_\sigma + \varepsilon_1 - \varepsilon_2)t} \\ 
    - 2i\sin{(\sqrt{2} g_\sigma - \varepsilon_1 + \varepsilon_2)t}
    \bigg|.
\end{gathered}
\end{equation}
Examples of the computed concurrence $C(t)$ for points $(p,q) = (0,1)$, $(p,q)=(1,2)$, and between them are presented in Fig.~\ref{fig:C-t}. In the intermediate region [Fig.~\ref{fig:C-t} (b, c, d)] $C(t)$ also reaches its maximum, which is, however, less than 1, and is achieved at slightly different time moments than in Fig.~\ref{fig:PQsets}. 
Interestingly, for the $q = p + 1$ subset, the maximal concurrence is the first local maximum among all local maximums of $C(t)$. Indeed, for any $0 < \phi < \pi/2$ concurrence $C=1$ will be observed at some time moment, but this point will be the part of a different subset with bigger $\tau_e$. For example, in Fig.~\ref{fig:C-t} (b) $C(t)$ becomes almost 1 at $t \approx 35$~ns, but this value belongs to higher branch with $q = p + 2$ and shall not be considered. Instead, the maximal concurrence is expected at $t \approx 9$~ns, and is evidently a shade less than 1. This definition of the maximal possible concurrence is justified by the entanglement reduction caused by the decoherence. When the damping and dephasing effects are taken into account, the much smaller gate time becomes way more important than a small correction to the concurrence that is expected for the pure state. 

Before concurrence reaches its maximum value (i.e. before the gate is performed), it oscillates at a frequency $\propto \phi$ (Fig.~\ref{fig:C-t}). Whereas $C(t)$ contains a number of local maxima, we should consider that one, which provides maximal entanglement together with shortest gate, as for the $q=p+1$ branch of interest. Specifically, we define the time moment, at which the gate is performed, as depicted in Fig.~\ref{fig:C-t} by dashed lines indicating the first local maximum that is bigger than the neighbouring peaks. Moving from the $(p,q)=(0,1)$ point [Fig.~\ref{fig:C-t} (a)] to higher $\phi$ values [Fig.~\ref{fig:C-t} (b)] significantly reduces the gate time, because the increase of spin-charge hybridization accelerates spin transitions and allows for faster energy transfer between the SQs. However, the reduction of the resulting entanglement is evident in Fig.~\ref{fig:C-t} (b), since the processes of photon emission and absorption by the first and second qubit do not interfere constructively, as opposed to Fig.~\ref{fig:C-t} (a). Increasing the value of $\phi$ makes the mismatch between the aforementioned processes more and more crucial and weakens the achieved entanglement. At a critical $\phi$ value, the concurrence becomes the same for two neighbouring peaks [Fig.~\ref{fig:C-t} (c)]. At this point, the resulting concurrence is minimal, whereas the gate time increases instantly by approximately $\frac{\pi}{\sqrt{2} g_\sigma}$. After the break, the peak entanglement grows when moving to stronger spin-charge coupling, as seen in Fig.~\ref{fig:C-t} (d). Meanwhile, the gate time goes down because of the same reason why it shrinks between Fig.~\ref{fig:C-t} (a) and Fig.~\ref{fig:C-t} (b). Finally, when $\phi$ reaches the $(p,q) = (1,2)$ point, the maximal concurrence returns to 1 [Fig.~\ref{fig:C-t} (e)]. 

Importantly, the entangling process in Fig.~\ref{fig:C-t} (e) is longer than in Fig.~\ref{fig:C-t} (a), as we already mentioned when discussing the lowest branch $q=p+1$ in Fig.~\ref{fig:PQsets}. We attribute this peculiarity to the fact that the rate of spin transitions and the trajectory of state rotation both depend on the degree of spin-charge hybridization characterized by $\phi$. On the one hand, big $\phi$ value enables SQs to exchange energy quickly, thereby reducing the entangling time. On the other hand, the mismatch between photon emission and absorption extends the trajectory of the SQs state rotation to the desired maximally entangled state. We reveal that within each branch presented in Fig.~\ref{fig:PQsets}, the latest factor always takes precedence. For example, at the $(p, q) = (8, 9)$ point with $\phi = 0.89 \frac{\pi}{2}$ a perfect entanglement is formed after 9 oscillations of the $\ket{\downarrow \downarrow}_0$ state population in $\tau_e = 20$~ns, as opposed to $\tau_e = 11$~ns and only one oscillation at the $(p, q) = (0, 1)$ point with $\phi = 0.097 \frac{\pi}{2}$. 

\begin{figure}
    \centering
    \includegraphics[width=0.9\linewidth]{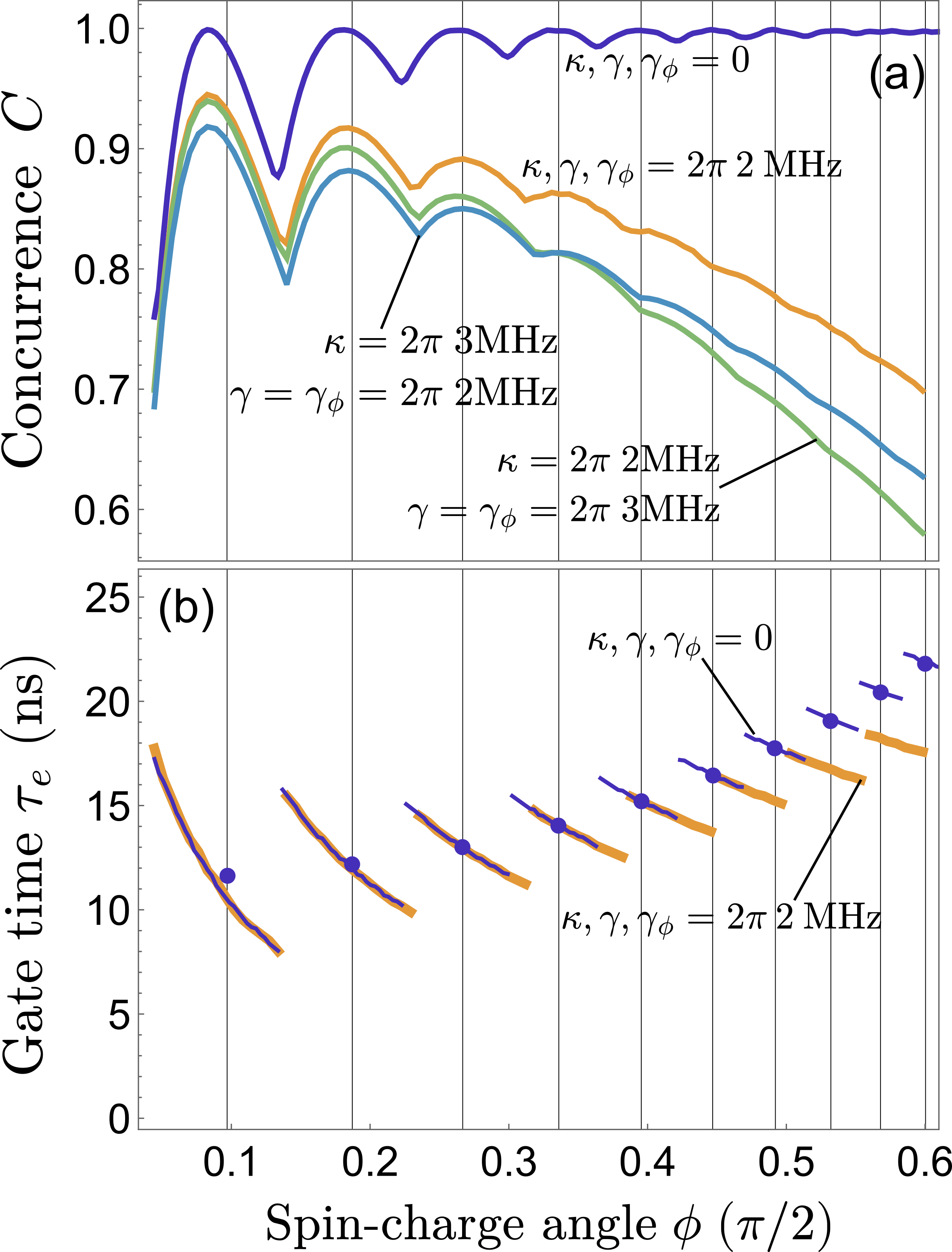}
    \caption{The concurrence $C$ (a) and gate time $\tau_e$ (b) as functions of angle $\phi$. Solid lines represent numerically derived data, while dots are taken from the $q=p+1$ branch in Fig.~\ref{fig:PQsets}. }
    \label{fig:F-tau-phi}
\end{figure}

To check how the entanglement given by the $q = p + 1$ branch is affected by decoherence, we calculate the achieved SQs concurrence $C(\phi)$ together with the gate time $\tau_e(\phi)$ at different $\phi$ values (note that from now on, $\tau_e(\phi)$ is considered to be a continuous function). To that end, the master equation~\eqref{eq:master} is solved numerically for the Hamiltonian from Eq.~\eqref{eq:H} with the Hilbert space dimension $4 \times 4 \times 3$, where $4 \times 4$ is the dimension of the SQs subspace, and $3$ is the artificially set upper limit of the cavity states. Note that the additional numerical calculations did not show any meaningful deviation of the derived results from that obtained at larger limit for the number of the cavity states. The result is presented in Fig.~\ref{fig:F-tau-phi} together with $q = p + 1$ branch from Fig.~\ref{fig:PQsets}.

Interestingly, even when neglecting decoherence, the time $\tau_e$, decreases monotonically but generally grows due to breaks located near the points of the minimal concurrence. The decrease of $\tau_e$ is intuitively clear: strong channel capacity results in high-speed gates. The breaks with suddenly increasing $\tau_e$ signify the change in trajectory between initial state and entangled state. To clarify why this is so, we appeal to Fig.~\ref{fig:C-t}, where such a break happens when we move from the panel (b) to the panel (c). In panels (a, b) the maximal concurrence is the second local maximum. When $\phi$ becomes bigger, the third local maximum approaches the second one. The break happens when these two maxima are of the same value, this $\phi$ also corresponds to the minimum of the concurrence in the $C(\phi)$ function [Fig.~\ref{fig:F-tau-phi} (a)]. After the trajectory is changed the achieved concurrence grows with $\phi$, while time to derive it goes down. 

In the same Fig.~\ref{fig:F-tau-phi} we also present the results for the mixed states computed by solving the master equation~\eqref{eq:master} numerically at different damping and dephasing rates. Importantly, the introduction of decoherence affects mostly the data situated at big spin-charge angles $\phi$, because the function $\tau_e(\phi)$ grows generally. Therefore, contra-intuitively, the ideal $\phi$ value appears to be quite small, around $9^\circ$ only at the chosen parameters. The small deviation of numerical data even at zero Lindbladian from our predictions at small $\phi$ can be attributed to the broken condition $g_\sigma \gg \frac{g_\tau^2}{\Delta}$, which should be met for the validity of the model. 

\begin{figure}
    \centering
    \includegraphics[width=0.93\linewidth]{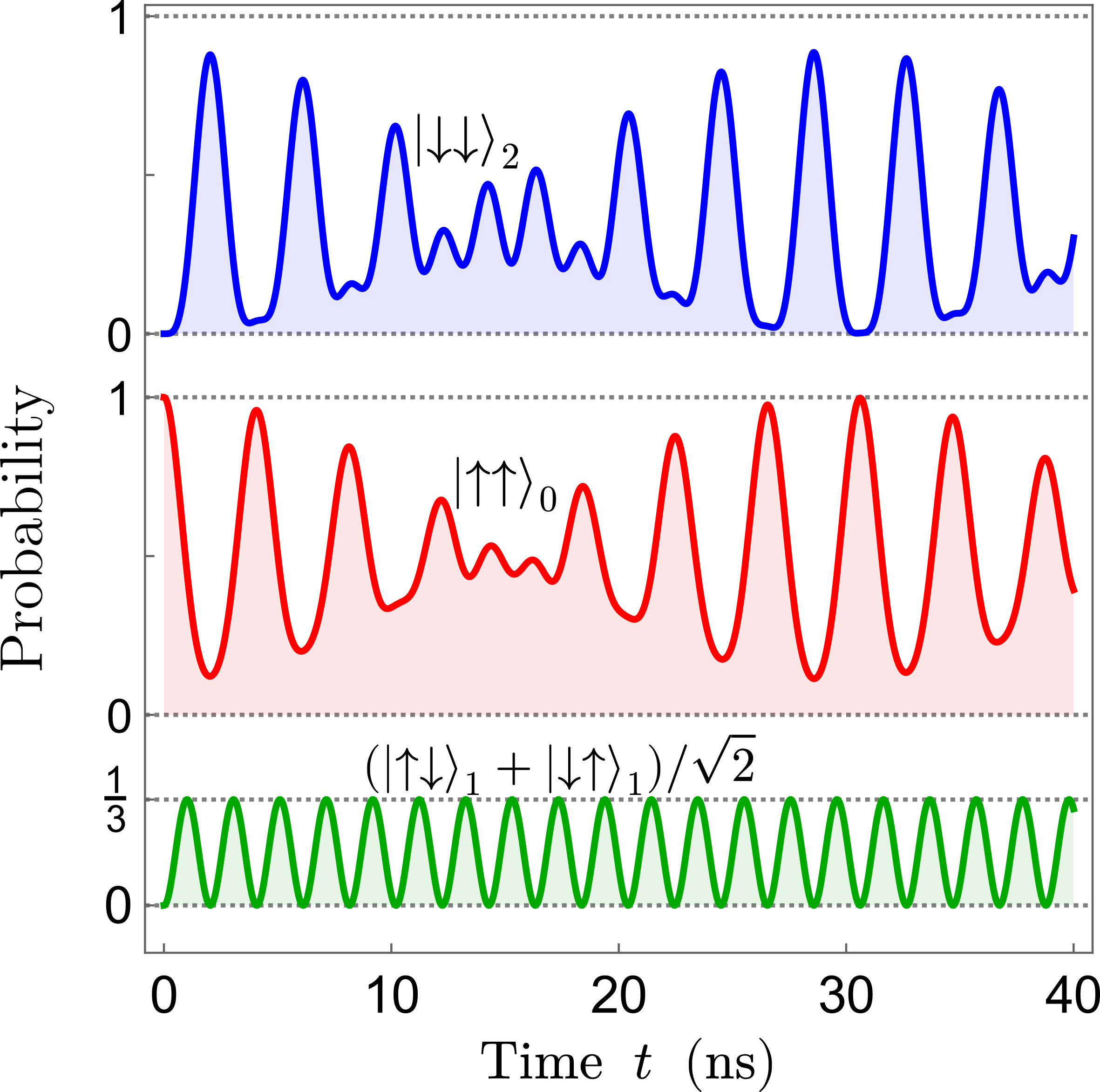}
    \caption{The temporal dependences of states probabilities for the initial condition $\ket{\Psi_0} = \ket{\uparrow\uparrow}_0$. The spin-charge angle is $\phi = \pi / 6$. }
    \label{fig:P-t_up-up}
\end{figure}

Next, we shortly discuss the case with initial condition $\ket{\Psi_0} = \ket{\uparrow \uparrow}_0$, which can be prepared via simultaneous application of two $\pi$-pulses to both SQs. In this case the wave function of the system is
\begin{widetext}
\begin{equation}
\begin{gathered}
    \ket{\Psi} = \frac{1}{6}\left( 4 \exp{[i \varepsilon_2 t]} + \exp{[i(\varepsilon_1 + \sqrt{6} g_\sigma)t]} + \exp{[i(\varepsilon_1 - \sqrt{6} g_\sigma)t]} \right) \ket{\uparrow\uparrow}_0 \\
    +\frac{1}{2 \sqrt{6}} \left( \exp{[i(\varepsilon_1 + \sqrt{6} g_\sigma)t]} - \exp{[i(\varepsilon_1 - \sqrt{6} g_\sigma)t]} \right) ( \ket{\uparrow\downarrow}_1 + \ket{\downarrow \uparrow}_1 )\\
    +\frac{1}{3 \sqrt{2}} \left( \exp{[i(\varepsilon_1 + \sqrt{6} g_\sigma)t]} + \exp{[i(\varepsilon_1 - \sqrt{6} g_\sigma)t]} -2\exp{[i \varepsilon_2 t]} \right) \ket{\downarrow\downarrow}_2,
\end{gathered}
\label{eq:solution:up-up}
\end{equation}
\end{widetext}
where $\varepsilon_1 = \Delta + \frac{3 g_\sigma^2}{2 \omega} + \frac{2 (7 \omega + 10 \Delta) g_\tau^2}{3\Delta(2 \omega + \Delta)}$ and $\varepsilon_2 = \Delta + \frac{g_\sigma^2}{\omega} + \frac{2 (7 \Delta + 4 \omega)g_\tau^2}{3 \Delta(2\omega+\Delta)}$. The corresponding probabilities are plotted in Fig.~\ref{fig:P-t_up-up}. The second term in Eq.~\eqref{eq:solution:up-up} represents an entangled state derived earlier, but with photon number $n = 1$. The maximal probability of this state is, however, only $1/3$. Thus, after tracing the cavity states out the obtained mixed state is characterized by relatively small concurrence between the two SQs. Thus, the initial condition $\ket{\Psi_0} = \ket{\uparrow\uparrow}_0$ cannot lead to any significant entanglement between the SQs and shall not be considered. Indeed, when the energy of the initial state $\ket{\Psi_0}$ exceeds zero, the excitations can be spread over a larger (than 2) number of cavity states, which leads to significant leakage of energy to higher cavity states with $n\geq2$. This can only reduce the probability of the desired maximally entangled state. 

\section{Conclusions \label{sec:conclusions}}
In conclusion, we have studied the process of entangling between the two SQs that are brought into resonant interaction with the shared superconducting microwave cavity. In order to describe the spin dynamics in the SQs, we performed a SW transformation to the first order and moved to a basis of spin-photon entangled states, diagonalizing the Hamiltonian. We predict that the Bell state $\frac{1}{\sqrt{2}}(\ket{\uparrow \downarrow}_0 - \ket{\downarrow \uparrow}_0)$ formation would take less than 8~ns from the initial state $\ket{\uparrow \downarrow}_0$, given an appropriate sequence of voltage pulses to detune the SQs from the resonator. We find that decoherence, primarily due to the cavity damping, reduces the concurrence to $C \approx 0.9$ for realistic parameter values. We also show that, for the free evolution of the combined three-partite system, the photonic energy oscillations represent $(i\mathrm{SWAP})^\alpha$ gates between the two SQs, where $\alpha$ can take large values, as opposed to the dispersive regime where $\alpha \ll 1$. This results in an unusual condition for the maximal entanglement between the two SQs from the $\ket{\uparrow \downarrow}_0$ initial state: the spin-charge hybridization must be relatively weak, $\phi \sim \frac{g}{\Delta}$. We reveal that the reason for that is the mismatch between the processes of emission and absorption of the real photon. As the interference of these processes becomes increasingly destructive when increasing the spin-charge angle $\phi$, a peculiar dependence of the entangling time on $\phi$ with breaks was discovered. 

Although putting the SQs into resonance with the cavity significantly reduces the entangling time, it also increases the fraction of energy stored by the electro-magnetic field. As a result, the cavity damping reduces the concurrence by at least $C_\mathrm{lost} \sim 0.1$. Notably, a similar value of the entanglement degree was predicted for the weakly coupled SQs in the dispersive regime, where $\sqrt{i\mathrm{SWAP}}$ entangling gate takes hundreds of nanoseconds. This is explained by much weaker decoherence in the dispersive limit, owing to the almost empty cavity. A key advantage of the described regime is that it significantly accelerates the entangling process, which cannot be implemented so fast with SQs that are detuned from the resonator. While the presented results capture the essence of the resonantly coupled dynamics of the three-partite quantum system, the most optimal conditions for implementing a set of universal gates remain unclear. The next important step in developing cavity-assisted gates for SQs is to identify the conditions that allow for fast and high-fidelity gates. 

\section{Acknowledgments}
This work has been supported by the Army Research Office
Grant No.~W911NF-23-1-0104.

\appendix
\section{Density matrix of the derived state \label{sec:appendix}}
Initializing the system in the $\ket{\Psi_0} = \ket{\uparrow \downarrow}_0$ state and keeping the ground-state SQ$_2$ detuned from the resonance with the cavity for time $\tau_1$, we assume that the second SQ is brought back into the resonance interaction, while the first SQ is now detuned. After time $\tau_2$ of the interaction between the second SQ and the cavity the system state is given by the following independent density matrix elements
  \begin{eqnarray}
      \prescript{}{0}{\bra{\downarrow \uparrow}} \rho \ket{\downarrow \uparrow}_0 &=& \frac{1}{4} \big[  \left( 1+e^{-\lambda \tau_1} (1-\cos{2 g_\sigma \tau_1}) \right) \nonumber \\
      & & \left( 1 - e^{-\lambda \tau_2} \cos{2 g_\sigma \tau_2} \right) \big], \nonumber \\
      \prescript{}{0}{\bra{\uparrow \downarrow}} \rho \ket{\uparrow \downarrow}_0 &=& \frac{1}{2}+e^{-\lambda \tau_1} \cos^2{g_\sigma \tau_1}, \nonumber \\
      \prescript{}{1}{\bra{\downarrow \downarrow}} \rho \ket{\downarrow \downarrow}_1 &=& \frac{1}{4} \big[  \left( 1+e^{-\lambda \tau_1} (1-\cos{2 g_\sigma \tau_1}) \right) \nonumber \\
      & & \left( 1 + e^{-\lambda \tau_2} \cos{2 g_\sigma \tau_2} \right) \big], \label{eq:rho-tau1+tau2} \\
      \prescript{}{0}{\bra{\downarrow \uparrow}} \rho \ket{\uparrow \downarrow}_0 &=& \frac{1}{2} e^{-(\lambda \tau_1 + \Lambda \tau_2)} \sin{2 g_\sigma \tau_1} \sin{g_\sigma \tau_2}, \nonumber \\
      \prescript{}{0}{\bra{\downarrow \uparrow}} \rho \ket{\downarrow \downarrow}_1 &=& \frac{i}{4} e^{-\lambda \tau_2} \left[ 1+e^{-\lambda \tau_1} (1 - \cos{2 g_\sigma \tau_1})  \right] \sin{2 g_\sigma \tau_2}, \nonumber \\
      \prescript{}{0}{\bra{\uparrow \downarrow}} \rho \ket{\downarrow \downarrow}_1 &=& \frac{i}{2} e^{-(\lambda \tau_1 + \Lambda \tau_2)} \sin{2 g_\sigma \tau_1} \cos{g_\sigma \tau_2} \nonumber,
  \end{eqnarray}
  where $\lambda = \frac{1}{4}(2\kappa + 2 \gamma_\phi \sin^2{\phi} + \gamma \sin^2{\phi})$ and $\Lambda = \frac{1}{4}(2 \kappa +6 \gamma_\phi \sin^2{\phi} + 3 \gamma \sin^2{\phi})$. 

\bibliography{ref.bib} 
\end{document}